%% file: paper.tex
\documentclass[10pt,conference]{IEEEtran}
\IEEEoverridecommandlockouts
\usepackage{cite}
\usepackage{amsmath,amssymb,amsfonts}
\usepackage{algcompatible}
\usepackage{algorithm}
\usepackage{algpseudocode}
\usepackage{mathtools}
\usepackage{graphicx}
\usepackage{textcomp}
\usepackage{xcolor}
\usepackage{balance}
\usepackage[all]{nowidow}
\usepackage[hidelinks]{hyperref}
\usepackage[noabbrev,capitalise]{cleveref}
\usepackage[caption=false]{subfig}

\usepackage{todonotes}
\newcommand{\tcnetem}{\text{NetEm}}

\def\BibTeX{{\rm B\kern-.05em{\sc i\kern-.025em b}\kern-.08em
    T\kern-.1667em\lower.7ex\hbox{E}\kern-.125emX}}
\begin{document}

\title{Network Emulation in Large-Scale Virtual Edge Testbeds: A Note of Caution and the Way Forward\\
    \thanks{Partially funded by the Deutsche Forschungsgemeinschaft (DFG, German Research Foundation) -- 415899119 and 414984028.}
}

\author{\IEEEauthorblockN{Soeren Becker\IEEEauthorrefmark{1}, Tobias Pfandzelter\IEEEauthorrefmark{2}, Nils Japke\IEEEauthorrefmark{2}, David Bermbach\IEEEauthorrefmark{2}, Odej Kao\IEEEauthorrefmark{1}}
    \IEEEauthorblockA{\IEEEauthorrefmark{1}\textit{Technische Universit\"at Berlin}\\
        \textit{Distributed and Operating Systems Research Group} \\
        \{soeren.becker,odej.kao\}@tu-berlin.de}
    \IEEEauthorblockA{\IEEEauthorrefmark{2}\textit{Technische Universit\"at Berlin \& Einstein Center Digital Future}\\
        \textit{Mobile Cloud Computing Research Group} \\
        \{tp,nj,db\}@mcc.tu-berlin.de}
}

\maketitle

\begin{abstract}
    The growing research and industry interest in the Internet of Things and the edge computing paradigm has increased the need for cost-efficient virtual testbeds for large-scale distributed applications.
    Researchers, students, and practitioners need to test and evaluate the interplay of hundreds or thousands of real software components and services connected with a realistic edge network without access to physical infrastructure.

    While advances in virtualization technologies have enabled parts of this, network emulation as a crucial part in the development of edge testbeds is lagging behind:
    As we show in this paper, \tcnetem{}, the current state-of-the-art network emulation tooling included in the Linux kernel, imposes prohibitive scalability limits.
    We quantify these limits, investigate possible causes, and present a way forward for network emulation in large-scale virtual edge testbeds based on eBPFs.
\end{abstract}


\input{sections/1_introduction}
\input{sections/2_background}
\input{sections/3_analysis}
\input{sections/4_solutions}
\input{sections/6_conclusion}

\balance

\bibliographystyle{IEEEtran}
\bibliography{bibliography}

\end{document}

%% file: sections/1_introduction.tex
\section{Introduction}
\label{sec:introduction}

As edge computing and the Internet of Things are becoming more important in both research and industry, the demand for environments to test new algorithms, platforms, and software systems is increasing~\cite{paper_bermbach_fog_vision,paper_pfandzelter_zero2fog}.
Between simulation, which can efficiently analyze an abstracted subset of an edge or IoT system~\cite{zeng2017iotsim,paper_hasenburg_fogexplorer,paper_hasenburg_fogexplorer_poster}, and physical infrastructure, which is accurate yet expensive to implement and cumbersome to maintain~\cite{csenel2021edgenet,wang2021tiansuan}, virtual testbeds running real software on virtualized infrastructure have emerged.
These testbeds enable a wide audience of researchers, students, and practitioners to evaluate software systems on large-scale distributed infrastructures by combining existing tooling for virtualization and emulation~\cite{paper_hasenburg_mockfog,paper_hasenburg_mockfog2,9020466}.

A key factor in building these testbeds for large-scale infrastructure is the scalability of the underlying tools itself.
Yet, while process isolation and machine virtualization have received significant interest from industry with their broad adoption in fields such as cloud computing~\cite{agache2020firecracker,paper_ernst_container_ecosystem}, the network emulators required to reflect the communication characteristics of edge and IoT are lagging behind.

The Linux traffic control network emulator (\tcnetem)~\cite{brown2006traffic} was originally introduced in kernel version 2.6.7 and has been continuously extended since then.
Because of its wide availability, maturity, and high accuracy~\cite{jurgelionis2011empirical,nussbaum2009comparative,lubke2014measuring,9020466}, \tcnetem{} is a popular choice for network emulation in edge and IoT testbeds, e.g.,~\cite{paper_hasenburg_mockfog,paper_hasenburg_mockfog2,9679447, paper_pfandzelter2022celestial,techreport_pfandzelter_celestial_extended,herrnleben2020iot,8898918,ahmed2017flexible,8958741,behnke2019hector}.
We find, however, that \tcnetem{} has only limited applicability for \emph{large}-scale testbeds as it struggles to support networks of hundreds or even thousands of nodes efficiently.
In this paper, we show these limits to enable future research into large-scale virtual edge and IoT testbeds and outline a possible way forward.

The rest of this paper is structured as follows:
We introduce \tcnetem{} and related concepts as well as alternative approaches in \cref{sec:background}.
In a scalability analysis using the context of virtual edge testbeds, we analyze the performance of \tcnetem{} at scale, exposing bottlenecks in configuration, latency overheads, and impact on network throughput (\cref{sec:analysis}).
We use novel Linux kernel features such as extended Berkeley Packet Filters (eBPF) and the EDT model to propose an alternative approach to network emulation for large scale IoT and edge testbeds and present promising preliminary performance results (\cref{sec:solutions}).
Finally, we conclude in \cref{sec:conclusion}.

%% file: sections/2_background.tex
\section{Background \& Related Work}
\label{sec:background}

The Linux traffic control subsystem (\texttt{tc}) provides mechanisms to control the transmission of packets in the Linux kernel~\cite{brown2006traffic}.
The main component of the traffic control subsystem are queuing disciplines (\texttt{qdisc}) that queue outgoing packets for a network device.
The default queuing discipline is a FIFO queue, but other strategies are available to support more fine-grained control over traffic shaping, e.g., by prioritizing packets that match a certain description.
This matching is performed by filters used to classify packets into queuing disciplines based on a classifier.

The \tcnetem{} network emulator (\texttt{tc-netem}) is an extension to the traffic control subsystem introduced to aid in the development of network protocols~\cite{hemminger2005network}, succeeding NIST Net~\cite{10.1145/956993.957007} and Dummynet~\cite{rizzo1997dummynet}.
At its core, \tcnetem{} is itself a queuing discipline, but provides further capabilities to emulate packet delay, loss, duplication, and reordering.
Several empirical evaluations have shown \tcnetem{} to be more accurate and to have better performance than competing network emulators~\cite{jurgelionis2011empirical,nussbaum2009comparative,lubke2014measuring,9020466}, yet have not considered scalability of the emulator.

Recent advances include extended Berkeley Packet Filters (eBPF)~\cite{ebpfspec} and eXpress Data Path (XDP)~\cite{10.1145/3281411.3281443}.
An evolution of the original Berkeley Packet Filters (BPF), eBPF programs run in a kernel virtual machine, are statically checked when loaded into the kernel, and are limited in size and complexity.
Applications can extend the kernel at runtime safely by attaching an eBPF program to one of the exposed hooks.
As the lowest level of the Linux networking stack, XDP is one such hook and supports processing packets by eBPF even before memory is allocated by the operating system~\cite{vieira2020fast}.

Based on XDP, Hemminger has presented XNetEm~\cite{hemminger2017xnetem} as an update to the original \tcnetem{}.
XNetEm runs an eBPF program as a kernel plugin that hooks into XDP processing on a network device and emulates packet loss, corruption, and marking.
XDP offers high performance for packet processing yet is limited in functionality and cannot be used to inject emulated network delays, which are also required for packet reordering.
As an alternative, the author suggests using Linux Traffic Control hooks that could expose the required functionality, an approach we take in our system (\cref{sec:solutions}).

Kumar et al.~\cite{10.1145/2785956.2787478} describe Bandwidth Enforcer (BwE), a solution to bandwidth allocation at Google.
After finding that existing approaches with policies enforced on network routers cannot support TCP traffic efficiently, the authors develop BwE to enforce service-specific limits on hosts.
In their work they leverage a global knowledge of the network topology in order to offer a hierachical bandwidth allocation for competing services across clusters of nodes connected in a WAN environment.

Saeed et al.~\cite{10.1145/3098822.3098852} show that this approach of host-based traffic shaping at scale consumes considerable CPU and memory resources on the hosts and present Carousel as an alternative.
In Carousel, a timestamp for each packet is computed before it is enqueued in a time-based queue and dequeued when the timestamp is met. By deploying a single queue per CPU core and enabling multicore lock-free coordination, Carousel is able to significantly increase the resource efficiency of host-based traffic shaping.

Although similar to our approach, BwE and Carousel aim to improve the network performance and utilization in wide area networks and data center communications whereas
we target network emulation.

%% file: sections/3_analysis.tex
\section{Scalability Analysis}
\label{sec:analysis}

Our performance analysis of \tcnetem{} concerns the overhead introduced by network emulation in the context of a virtual edge testbed.
In this section, we describe this scenario in more detail (\cref{subsec:analysis:scenario}), show bottlenecks encountered during \tcnetem{} configuration (\cref{subsec:analysis:setup}), and consider latency (\cref{subsec:analysis:latency}) and bandwidth overheads (\cref{subsec:analysis:bandwidth}).
We make our software artifacts available as open-source\footnote{\url{https://github.com/srnbckr/ebpf-network-emulation}}.

\subsection{Motivation and Scenario}
\label{subsec:analysis:scenario}

Our experiments are motivated by our work on Celestial~\cite{paper_pfandzelter2022celestial,techreport_pfandzelter_celestial_extended} which emulates low earth orbit (LEO) edge infrastructure such as SpaceX' Starlink constellation.
In Celestial, individual satellites are emulated through a Firecracker microVM~\cite{agache2020firecracker} each and link characteristics are shaped through \tcnetem{}.
Due to the sheer number of satellites in a constellation (which is usually in the 5,000 to 10,000 range according to current plans), the scalability of \tcnetem{} is a significant factor.
Therefore, we derive the scenario characteristics from the Celestial use case and build on the Celestial prototype for experiments, leading to the following scenario:

We assume a fully meshed topology of $N$ processes, each with their own network address.
We assume all processes running on a single host.
With emulated connections between these processes that let us make individual changes to the network link characteristics, we require emulation of $N \times (N - 1)$ network links.
This assumes one link per direction, e.g., a link attached to process A emulating network characteristics to a process B and a further link for the opposite direction attached to process B.
While this does not reflect all network topologies encountered in IoT or edge computing deployments, it serves as a straightforward baseline example and allows us to show scalability in relation to the factor $N$.

Per process, we create a \texttt{tap} network device.
All applications used for performance measurements are run in Firecracker microVMs attached to one of the network devices.
We configure a root hierarchical token bucket (HTB) queuing discipline for each network device.
For each outgoing link from this network device, we add a subordinate HTB queuing discipline using an ascending index.
We then set the desired \tcnetem{} emulation properties on this queuing discipline and attach a filter that matches by destination IP address.
Our experiments are executed on a single Google Cloud Platform \texttt{n2-standard-32} machine with 32~vCPUs and 128~GB memory running Ubuntu 20.04 LTS in the \texttt{eu-west-3} region.

\subsection{Configuration Bottlenecks}
\label{subsec:analysis:setup}

\begin{figure}
    \centering
    \includegraphics[width=\linewidth]{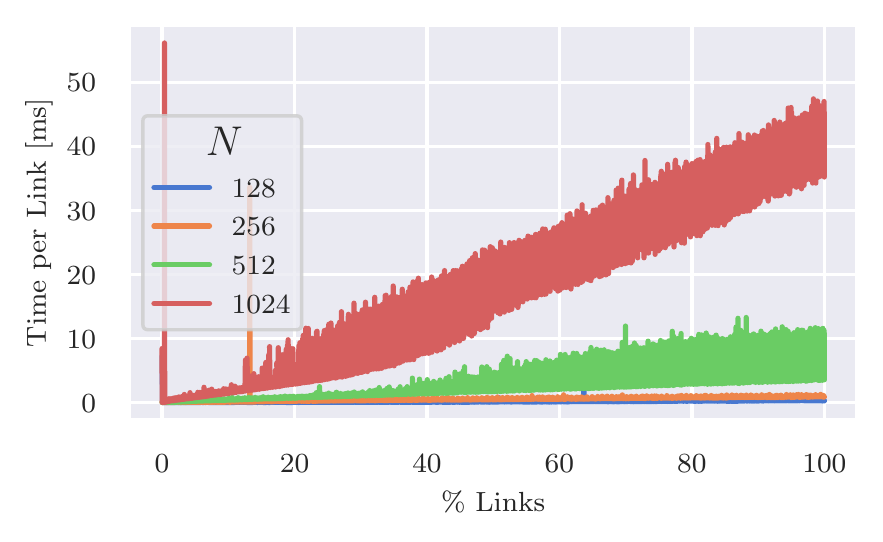}
    \caption{As the percentage of configured links grows during the \tcnetem{} configuration, the time per configuration change increases linearly. At $N = 1024$, configuring one link takes up to 47.4ms.}
    \label{fig:timeperlink2}
\end{figure}

Before examining the network performance impact of \tcnetem{} at scale, we consider the overhead of actually configuring link emulation.
We observe considerable bottlenecks that inhibit large-scale emulated testbeds.

We consider $N$ processes and $N \times (N - 1)$ emulated links.
In our experiment, configuring link emulation for the first link takes only 50\textmu s.
At the order of $N = 1000$, this overhead would accumulate to 50s if links were configured sequentially.
Yet as \cref{fig:timeperlink2} shows, the overhead of configuring one link grows with the number of links that are already configured.
As a result, the time to create links with $N = 1024$ grows to 19,422.7ms, or three hours and 24 minutes.

We identify two possible causes for this behavior:
First, new queuing disciplines are attached to the root discipline in a tree, which requires traversing existing handles when attaching a new one.
Second, the kernel might check for duplicate or otherwise conflicting queuing discipline entries, requiring reading all existing entries.

In a further test, we try to parallelize the filter creation by configuring each network device with a separate process.
In theory, these links have independent queuing disciplines, and we expect a significant speed-up in link configuration.
Unfortunately this did not change our results, possibly because a global lock for Traffic Control prevents concurrent changes to the network subsystem.

We decided to not further investigate and remedy the causes of the effects we observe in setting up large numbers of links for \tcnetem{} as our performance measurements do not justify using \tcnetem{} at this scale.
Further, we note that our alternative approach (\cref{sec:solutions}) does not exhibit any such configuration bottlenecks.

\subsection{Latency}
\label{subsec:analysis:latency}

\begin{figure}
    \centering
    \includegraphics{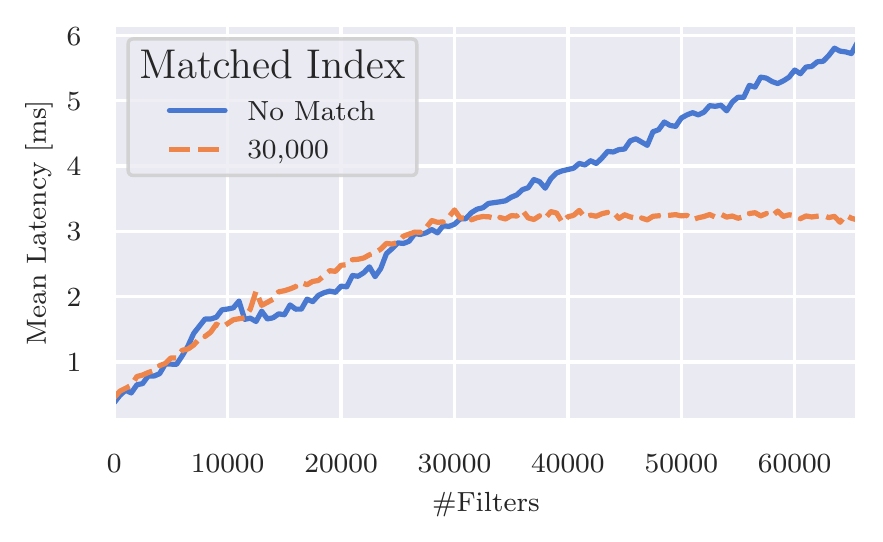}
    \caption{Experiment results show that network delay increases as more filters and queuing disciplines are attached to the device. The latency overhead of \tcnetem{} is only approximately 0.1\textmu s per filter, yet this accumulates to 5.7ms with more existing filters. Using a filter that matches our target address at the 30,000th location reveals that filters are checked in sequence and a match limits that overhead.}
    \label{fig:latency}
\end{figure}

We first consider the network latency overhead caused by \tcnetem{}.
Using the setup described in \cref{subsec:analysis:scenario}, we run \texttt{ping} in one microVM and probe network delay to the underlying host.
We consider a growing number of links configured for this node, yet set emulated network delay to 0ms and disable bandwidth throttling in HTB in order to have results reflect only additional overhead caused by processing in \tcnetem{}.
We use a maximum of 65,000 links as we are limited by the $2^{16}$ address space of Traffic Control child handles.
To account for the impact of a filter matching our ICMP packets, processing it, and returning before additional filters are checked, we test two scenarios:
First, none of the filters match the target IP address, requiring \tcnetem{} to check all existing filters.
Second, we add a filter matching the host's address at the 30,000th index to see whether this earlier entry in the filter sequence leads to lower overheads when \tcnetem{} does not check remaining filters.

We show the results of our delay measurement in \cref{fig:latency}.
Since the connection between VM and host machine is virtual, there should be minimal networking overhead during performance measurements, and we measure an average of 0.3ms round trip time as a baseline.
With increasing numbers of filters attached to our network device, however, latency increases linearly to up to 6ms when processing 65,000 filters.
As expected, a filter matching our test packet means that no further filters are tested by \tcnetem{}.
In our tests, increasing the number of filters beyond 30,000 does not further increase latency of around 3.3ms when a filter matches at this index.

\subsection{Bandwidth}
\label{subsec:analysis:bandwidth}

\begin{figure}
    \centering
    \includegraphics{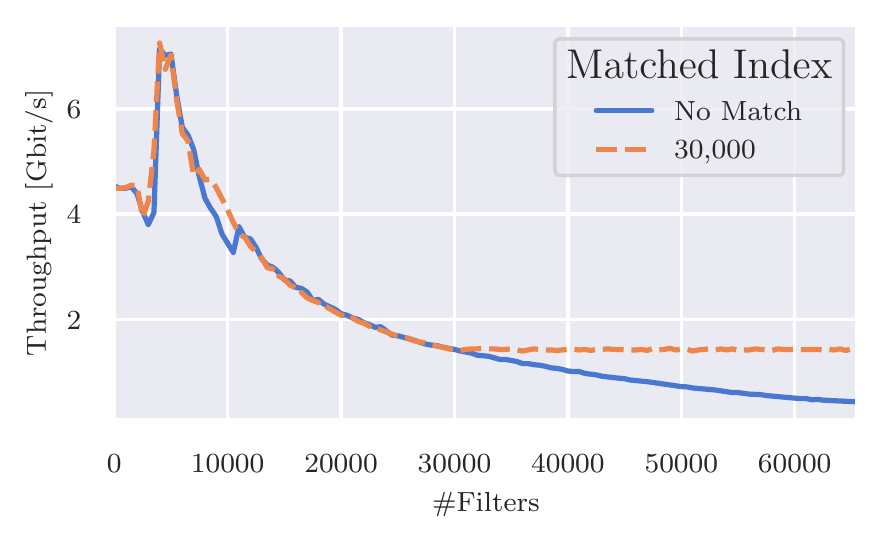}
    \caption{Throughput tests show a 4.6Gbit/s baseline without any filters yet up to 7.3Gbit/s using 5,000 filters. After this peak, throughput decreases steadily to only 0.4Gbit/s at 65,000. Again, matching filter 30,000 with our test packets limits the throughput overhead for experiments with additional filters.}
    \label{fig:bandwidth}
\end{figure}

Our evaluation of the bandwidth overhead of \tcnetem{} follows the same methodology as described in \cref{subsec:analysis:latency}, yet we use \texttt{iperf3}~\cite{iperf3} to perform TCP bandwidth measurements.
Again we measure once without any filter matching our test packets and once with the filter at index 30,000 aligning with the host's address.

The results of our tests with \tcnetem{} are shown in \cref{fig:bandwidth}.
As a baseline, we see a throughput of 4.6GBit/s without the use of any filters.
Surprisingly, we observe a small drop in throughput to 3.9Gbit/s when attaching 2,000 filters yet see it increase to 7.3Gbit/s with 5,000 filters.
While more throughput might be agreeable in general, unexpected effects of running \tcnetem{} at scale are undesirable in research testbeds.
In our case, we assume that mainly the interplay of nested virtualization with Firecracker, \texttt{tap} devices and \texttt{iperf3} induces this behaviour,
since it was not reproducable on a bare-metal node.
After this peak, throughput decreases steadily to only 0.4Gbit/s at 65,000 attached filters.
As in our latency measurements, we see how a filter matching our packets impacts those results, with an increase in filters beyond 30,000 not further reducing throughput in these tests.

%% file: sections/4_solutions.tex
\section{Solution}
\label{sec:solutions}

\begin{figure*}
    \centering
    \includegraphics[width=1\textwidth]{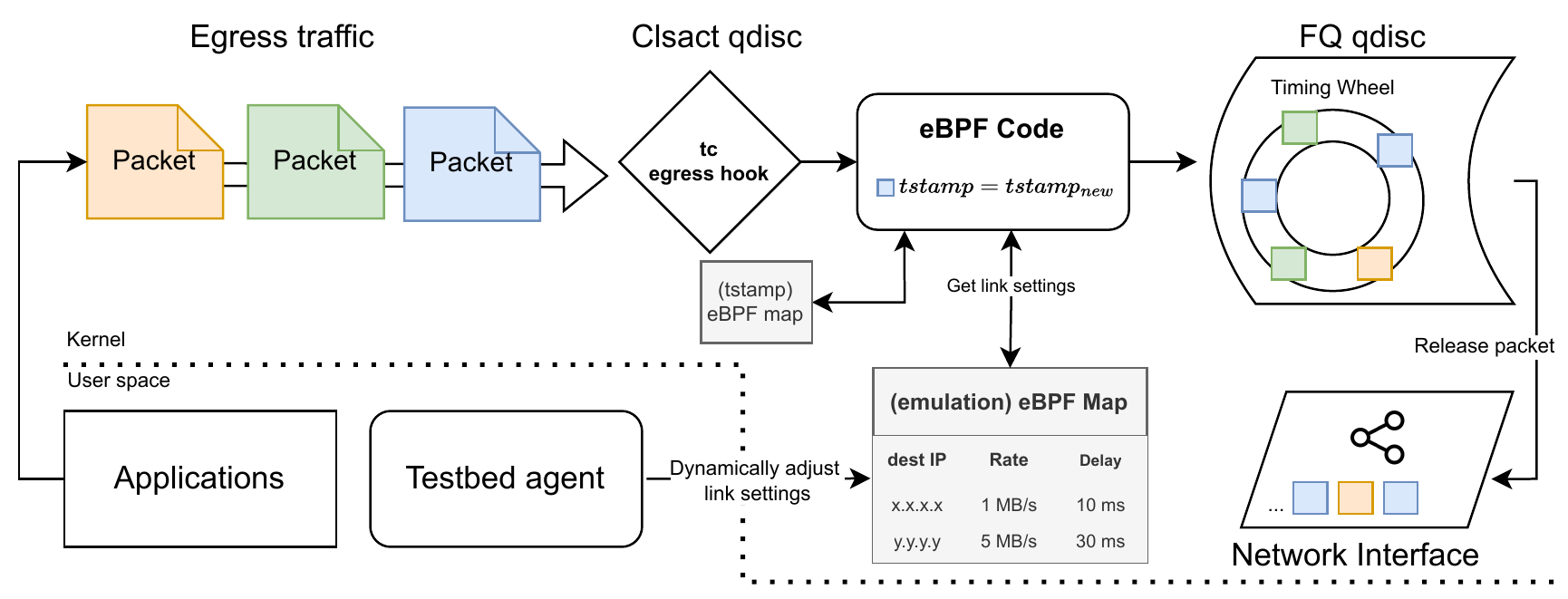}
    \caption{Our proposed solution relies on a \texttt{tc egress} hook that forwards outgoing packets to an eBPF program which adjusts the departure time of packets. The eBPF code obtains network emulation parameters used to compute the new timestamp from an eBPF map that is configured via a testbed agent and a timing wheel scheduler is employed to release packets to the network interface when the departure time is met. }
    \label{fig:approach}
\end{figure*}

In order to tackle the aforementioned scalability issues, we propose a network emulation approach that leverages a Traffic Control \texttt{classifier} and eBPF program to adjust the departure timestamp of outgoing packets and subsequently forwards them to a single fair queuing \texttt{(FQ)} queuing discipline.
A similar method was outlined in Carousel~\cite{10.1145/3098822.3098852}, yet only implemented in Software NICs since necessary features were not yet available in the kernel.
With the recent switch of the Linux networking stack to the early departure time (EDT) model, each egress packet has a departure timestamp which in turn can be adjusted with eBPF programs~\cite{hbtedt}.
Consequently, this enables the use of a timing wheel~\cite{vargheseHashed} packet scheduler that releases a packet to the network interface based on its departure timestamp instead of, e.g., \texttt{bfifo} or \texttt{pfifo} queues.
In recent works~\cite{10.1145/3098822.3098852,hbtedt} this model was used for rate limiting and pacing in, e.g., data centers.
We also consider it useful to efficiently emulate network characteristics in large scale virtual edge testbeds.
Therefore, we propose an approach as depicted in \cref{fig:approach} using the following components:

\begin{itemize}
    \item An eBPF map shared between the kernel and user space that contains network link parameters such as bandwidth limitations or latency to different destination IP addresses.
    \item A testbed agent running in user space which adjusts the map with parameters used to emulate the link settings to different nodes based on the destination IPs.
\end{itemize}

In addition, we attach an eBPF program to the locally generated traffic utilizing a Traffic Control \texttt{egress} hook which conducts the following steps for each outgoing packet:
Initially, a map lookup with the destination IP of the packet is executed to check the shared eBPF map for potential network emulation parameters.
In case no parameters exist for the destination, the packet is instantly forwarded to the network device, therefore only imposing a negligible overhead on traffic which should not be affected by any network emulation.
When parameters are found, the program computes a new timestamp for the given packet.

Here we propose a two-fold approach:
For a defined bandwidth limitation we apply an algorithm as outlined by Fomichev et al.~\cite{hbtedt}, that introduces rate-limiting by adjusting the inter-packet gaps, while utilizing the last timestamp of a packet obtained from a second eBPF map.
As described by \cref{alg:set_delay}, if a previous timestamp exists, the inter-packet gap is increased by a delay depending on the packet size and given rate.
Finally, the map is updated with the newly computed timestamp for the next packet.
We further extend this method by storing the latest packet timestamps for several IP addresses in order to enable varying network emulation settings based on the destination.

\begin{algorithm}
    \caption{Set the departure timestamp for egress packets}
    \label{alg:set_delay}
    \begin{algorithmic}[1]
        \State \textbf{Input:} $packet$
        \State $ip \gets destIP(packet)$
        \State $rate, latency \gets$ mapLookup($map_{emulation}, ip$)
        \If{rate}
        \State throttle($packet, rate$)
        \EndIf
        \If{latency}
        \State injectDelay($packet, latency)$
        \EndIf
        \State \textbf{return} $packet$

        \Procedure{Throttle}{$packet, rate$}
        \State $t_{last} \gets$ mapLookup($map_{tstamp}, ip$)
        \State $gap \gets  packet_{len} * 10^9 / rate$
        \If{$t_{last}$}
        \State $t_{next}\gets t_{last} + gap$
        \Else
        \State $t_{next} \gets now$
        \EndIf
        \State mapUpdate($map_{tstamp}, ip, t_{next}$)
        \State updateTimestamp($packet, t_{next}$)
        \EndProcedure
        \Procedure{InjectDelay}{$packet, rate$}
        \State $t_{next} \gets packet_{tstamp} + latency$
        \State updateTimestamp($packet, t_{next}$)
        \EndProcedure
    \end{algorithmic}
\end{algorithm}

Moreover, in case an additional latency to the destination was defined in the eBPF map, the timestamp of the packet is further increased by that delay.
Finally, the packet is enqueued in the timing wheel scheduler which subsequently releases the packet to the network interface when the departure timestamp is met.

Compared to \tcnetem{}, where in the worst case it is required to create a \texttt{filter} as well as HTB and \tcnetem{} queuing disciplines for each link that should be emulated, our approach relies on a single Traffic Control \texttt{classifier}, \texttt{FQ} queuing discipline, and shared eBPF map which significantly simplifies and accelerates the setup and adjustment process:
In our experiments, filling the eBPF map with 65,000 IP and network emulation parameter pairs takes around 170ms and the values can be updated with a simple user space program.
This is especially important considering the scenario depicted in \cref{subsec:analysis:scenario}, which requires a vast amount of network characteristics to be set for different links.
Additionally, IP pairs can be used as map keys, which allows to employ a single eBPF map describing all possible links.

We implement our approach as a proof-of-concept prototype that is able to limit bandwidth and inject latencies following the method outline above.
While we omit more sophisticated network emulation settings such as delay distribution, packet loss or corruption in this prototype, these could easily be implemented by adapting methods from XNetEm~\cite{hemminger2017xnetem}.

\paragraph*{Preliminary evaluation}

We conduct a preliminary evaluation of our prototype using the same experimental setup as described in \cref{subsec:analysis:scenario}.
Here, we employ the aforementioned \tcnetem{} method for up to 65,000 filters on a single network interface and compare the overhead to our eBPF approach and a map filled with the same amount of entries, repeating the experiments from \cref{subsec:analysis:latency} and \cref{subsec:analysis:bandwidth}.

\begin{figure}
    \centering
    \includegraphics[width=\linewidth]{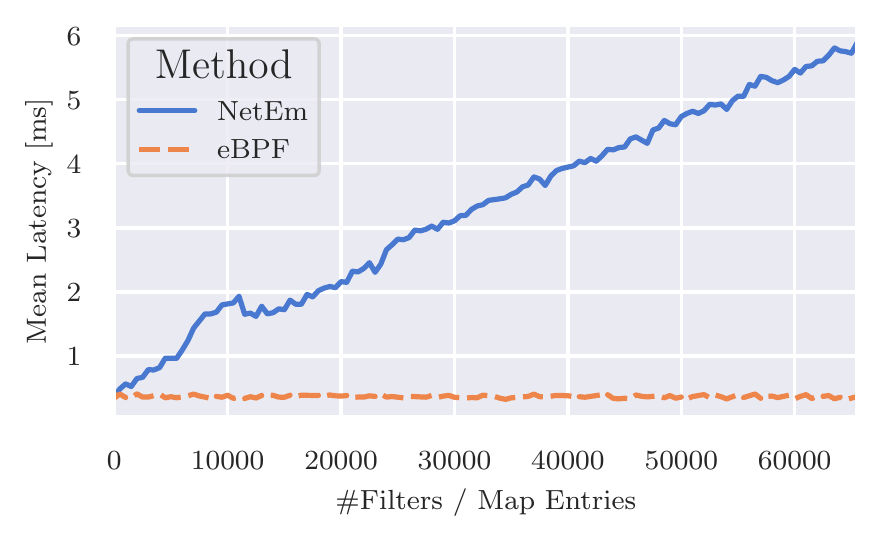}
    \caption{Overhead of increasing queuing disciplines and filters for \tcnetem{} as well as map entries for the eBPF approach on the latency, without any matches. For eBPF the overhead stays relatively constant.}
    \label{fig:newresults:latency}
\end{figure}

\cref{fig:newresults:latency} shows the increase in network latency overhead for ascending amounts of non-matching filter and map entries.
Whereas \tcnetem{} increases the latency to up to 6ms, our prototype only poses a constant and negligible overhead of around 0.1ms on the latency.
This is mainly due to the fact that packets can be checked in $\mathcal{O}(1)$ using our eBPF map whereas \tcnetem{} checks each filter individually in $\mathcal{O}(N)$.
In our prototype, non-matching packets are then directly released to the network interface.

\begin{figure}
    \centering
    \includegraphics[width=\linewidth]{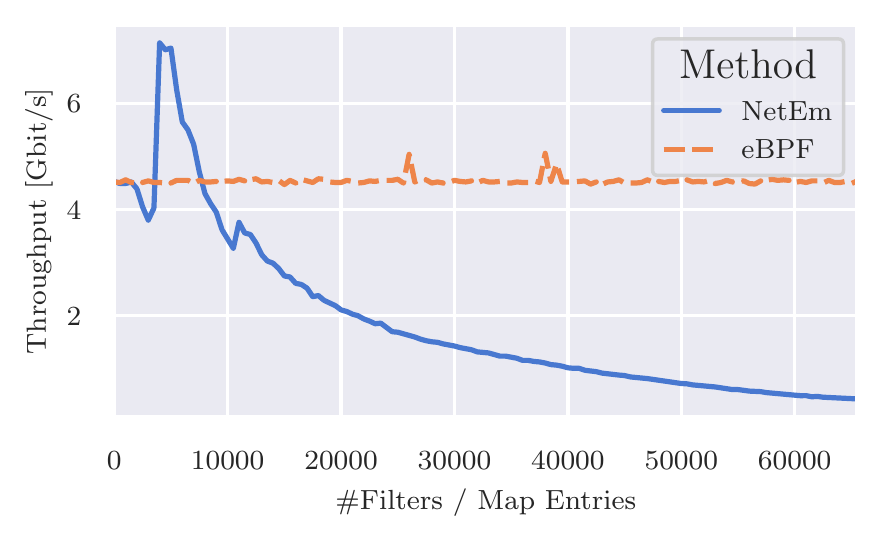}
    \caption{Overhead of increasing queuing disciplines and filters for \tcnetem{} as well as map entries for the eBPF approach on the bandwidth, without any matches. Non-matching entries have a negligible impact on the overall bandwidth in our prototype and it is not affected by an unexpected peak like \tcnetem{}.}
    \label{fig:newresults:bandwidth}
\end{figure}

\begin{figure}
    \centering
    \includegraphics[width=\linewidth]{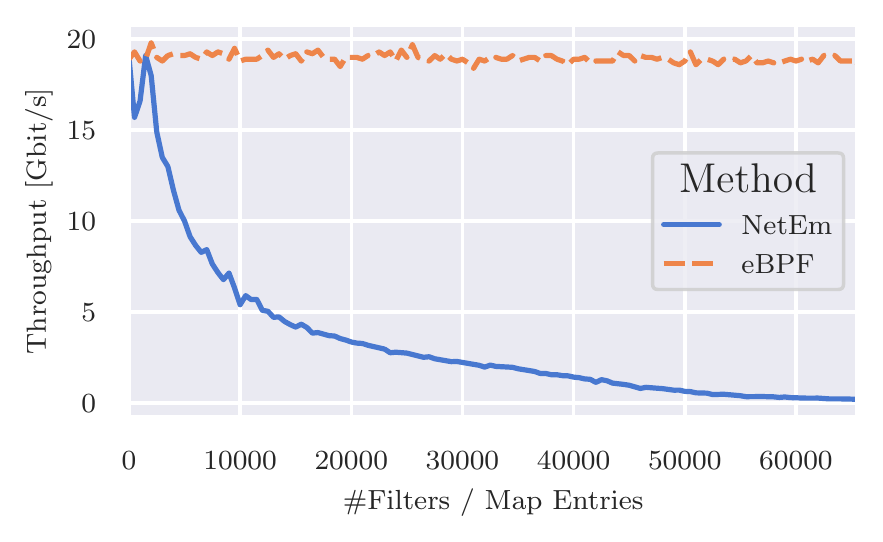}
    \caption{Bandwidth throughput overhead on a bare-metal machine for increasing queuing disciplines and filters or map entries for the \tcnetem{} respectively eBPF approach. Compared to nested virtualization, the baseline throughput is not exceeded anymore in case of \tcnetem{}, although the drop at around 2,000 filters remains. The overhead of the eBPF approach is still negligible.}
    \label{fig:newresults:bandwidth:metal}
\end{figure}

\begin{figure}
    \centering
    \includegraphics[width=\linewidth]{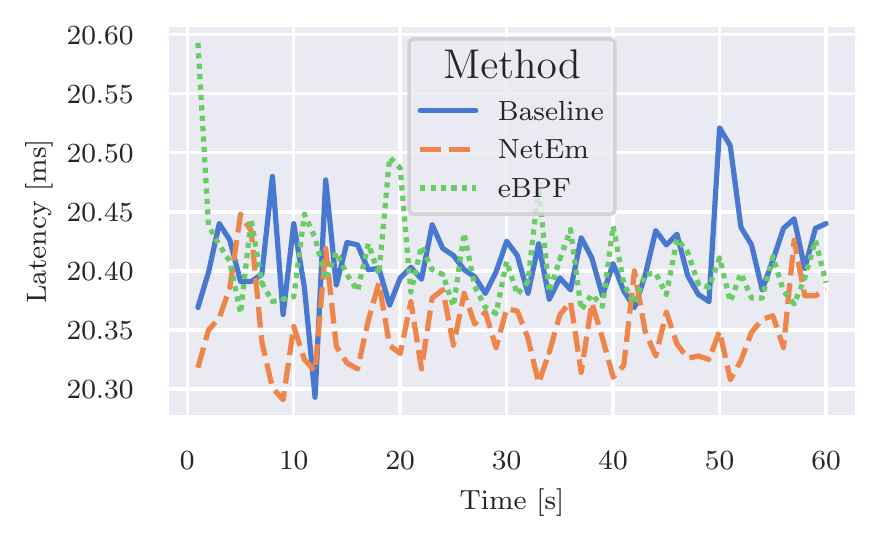}
    \caption{\texttt{Ping} results for an emulated delay of 20ms using the \tcnetem{} method and our approach. Baseline refers to the measured latency without any emulation, increased by 20ms.}
    \label{fig:latency_experiment}
\end{figure}

\begin{figure}
    \centering
    \includegraphics[width=\linewidth]{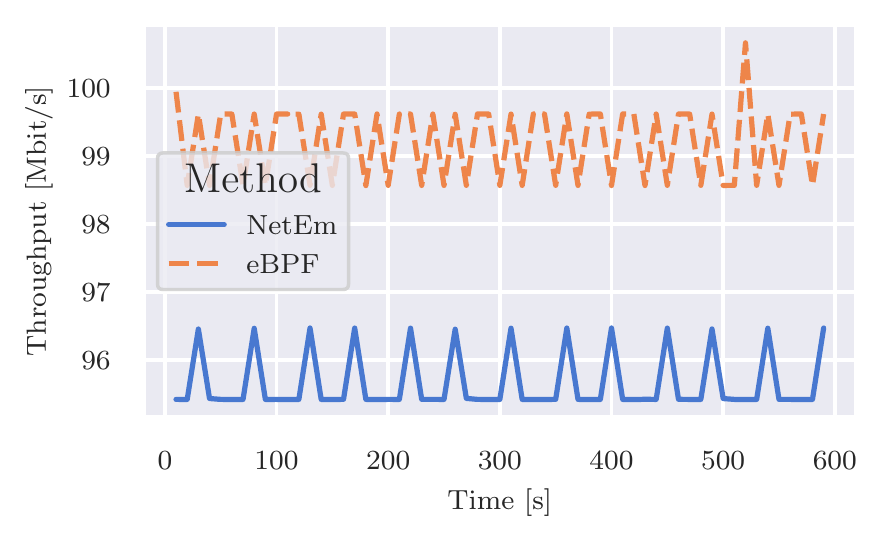}
    \caption{\texttt{iperf3} results for a bandwidth limitation of 100Mbit/s and emulated delay of 5ms. We show the mean throughput in the last 10 seconds for a duration of 10 minutes.}
    \label{fig:iperf_experiment}
\end{figure}

The same trend can be identified in \cref{fig:newresults:bandwidth}, which depicts the bandwidth overhead for non-matching filters: In case of our prototype,
the bandwidth throughput is around 4.53GBit/s and therefore close to the optimum, regardless of the amount of map entries.
As discussed in \cref{subsec:analysis:bandwidth}, we assume that nested virtualization with Firecracker microVMs results in the increase of
throughput at around 5,000 filters for \tcnetem{} and additionally tested the two methods on a \texttt{AMD EPYC 7282} bare-metal machine with 32 CPU cores.
Although the experiment was again conducted in two microVMs, the results depicted in \cref{fig:newresults:bandwidth:metal} show a baseline throughput of around 19.5GBit/s that is not exceeded anymore
at 5,000 filters, therefore supporting this assumption and motivating further research in scenarios deviating from the use case described 
in \cref{subsec:analysis:scenario}, that we mainly evaluated in this paper.

Finally, we also compared the accuracy of the latency injection and bandwidth limitation for both network emulation methods: \cref{fig:latency_experiment} shows the latency for an artificial delay
of 20ms, monitored for one minute using \texttt{ping} and added as a single filter respectively map entry. As a baseline we monitored the latency without any emulation and increased the values by 20ms.
As can be seen, our prototype is able to produce an emulation equivalent to \tcnetem{}.

Moreover, we also deployed two Firecracker microVMs, each connected to a \texttt{tap} device as described in \cref{subsec:analysis:scenario}, limited the bandwidth between the nodes to 100Mbit/s and added a latency of 5ms with a single filter and map entry. We tested the available throughput between the nodes with \texttt{iperf3}, running in the microVMs.
As can be seen in \cref{fig:newresults:bandwidth}, with both methods the throughput approaches the theoretical maximum and the eBPF emulation even slightly outperforms \tcnetem{} in terms of rate limit utilization in our experiment.

%% file: sections/6_conclusion.tex
\section{Conclusion \& Future Work}
\label{sec:conclusion}

In this paper, we discussed the caveats of using \tcnetem{} for building large-scale virtual testbeds for the edge and IoT.
In a scalability analysis, we showed that \tcnetem{} causes considerable performance overheads during setup and execution which make it unsuitable for larger emulated networks.

Using novel Linux kernel features such as eBPFs and the EDT model, we are able to remedy some of these scalability issues without relying on complex dependencies.
Our evaluation of this approach shows negligible latency and bandwidth overheads even with as many as 65,000 filters attached to a network device thanks to packet matching in constant time.
We are already integrating this solution into a large-scale virtual testbed and hope that it can benefit the research community in building the next generation of IoT and edge computing testbed tools.